# Special Relativity properties from Minkowski diagrams

**Nilton Penha[1] and Bernhard Rothenstein[2]**

[1] *Departamento de Física, Universidade Federal de Minas Gerais, Brazil* - nilton.penha@gmail.com.
[2] *Politehnica University of Timisoara, Physics Department, Timisoara, Romania* – brothenstein@gmail.com.

## Abstract

This paper has pedagogical motivation. It is not uncommon that students have great difficulty in accepting the new concepts of standard special relativity, since these seem contrary to common sense. Experience shows that geometrical or graphically exposition of the basic ideas of relativity theory improves student's understanding of the algebraic expressions of the theory. What we suggest here may complement standard textbook approaches.

## I - Spacetime Diagram

The two pillars of Einstein's special relativity theory are the following two postulates [1]:

1) All laws of physics are the same in all inertial frames;

2) In empty space, light travels with constant speed *c* independent of the source or the observer motions.

Any physical phenomenon takes place at a space position and at a certain time. To characterize this "event" we need to know the spatial and time coordinates in a four dimensional spacetime, as one usually refers to the set of all possible events. In such spacetime, an event is a point with coordinates $(x,y,z,t)$ or as most conventionally used $(x,y,z,ct)$, the *c* being the absolute value of light speed in free space*.

One given event has different coordinates depending on the reference system one is adopting. Special relativity deals with the transformation of the set of coordinates of any event in a given inertial frame into another set in any other inertial frame. The observers can measure space distances with measuring-rods and time with measuring-clocks. Both rods and clocks are assumed to be in all respects alike.

Minkowski spacetime diagram[2] is a graphical representation of events and sequences of events in spacetime as "seen" by observer at rest. Such sequences are named wordlines. A two-dimensional spacetime example is given in Figure 1, where $x$ and $ct$ are the two coordinates. A light ray emitted at the origin along the *x-axis* towards positive values of $x$ in space is represented by a straight line $ct = x$ (*OW*). You should notice that *OW* bisects the quadrant formed by *ct-axis* and *x-axis*); *GH* ($ct = -x$) represents a light ray emitted at the origin traveling along the *x-axis* towards negative values of $x$; *OP* represents anything traveling with moderate constant speed as compared to $c$ ; the slope of the worldline gives a measure of the speed; *OQ* represents anything traveling with constant speed close to that of light; *IJ* represents events happening simultaneously; The *ct-axis* itself represents anything standing still at point *O* $(x,ct) =$

---

* The exact speed of light, in vacuum, is 2.99792458 x $10^8$ km/s.

*(0,ct)*; the *x-axis* represents simultaneous events at *(x,ct)=(x,0)* (you may understand them as occurring "now"). The light worldline is called lightline. The locus of all lightlines starting at the origin constitutes a light cone, as shown in Figure 2 for three-dimensional spacetime. In two dimensions, as shown in Figure 1, *GOW* constitutes a light "cone". Although real physical phenomenon happens in four dimensional spacetime, one usually stick with the two-dimensional representation because most of what can be deduced from Minkowski diagrams can be done in this simple scenario.

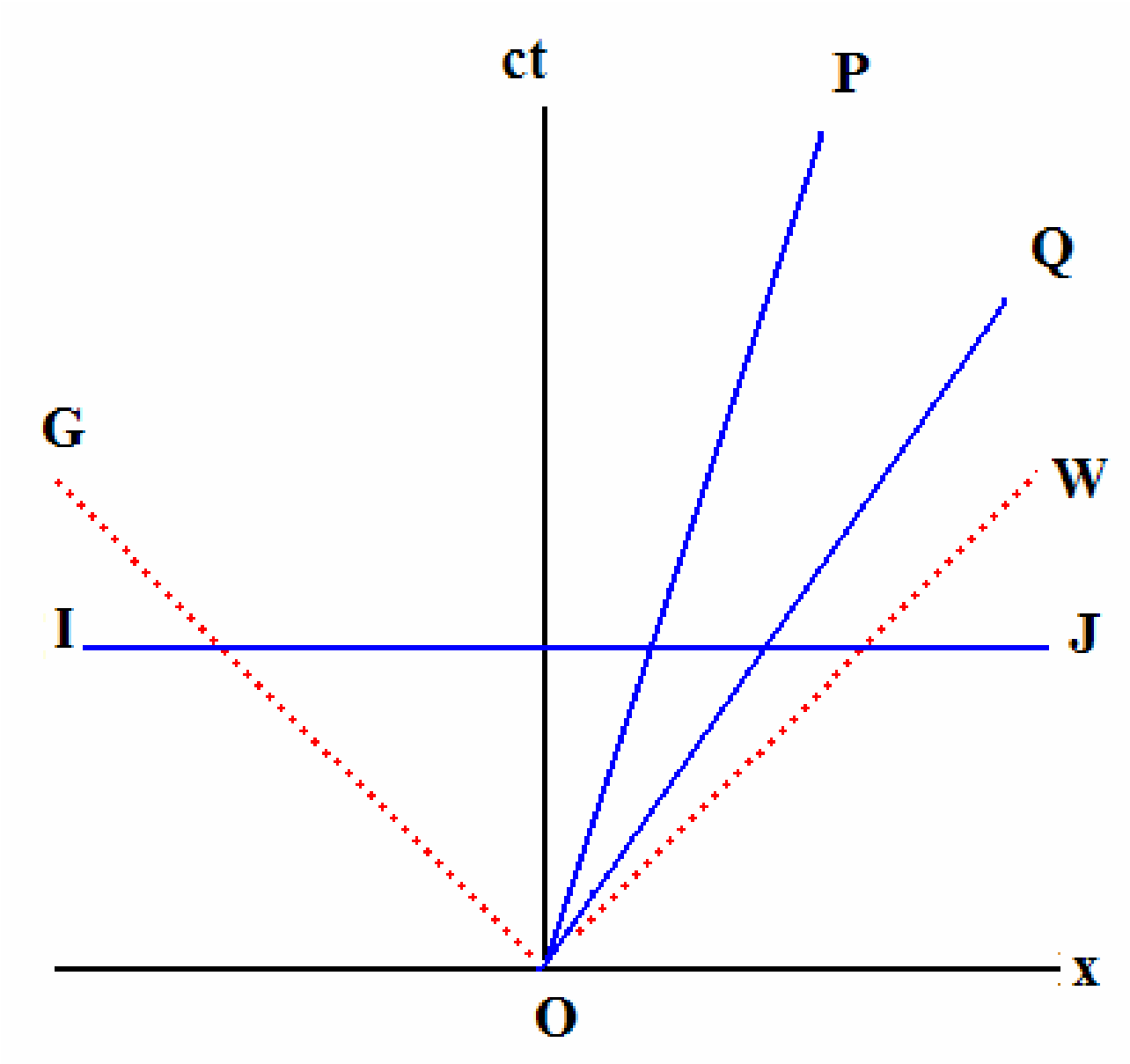

**Figure 1** – This shows spacetime diagram in two dimensions (*x,ct*). Worldline *OW* represents a light ray traveling along the *x-axis* towards positive values of *x* (notice that *OW* bisects the quadrant formed by ct-axis and *x-axis*); *GO* represents a light ray traveling along the *x-axis* towards negative values of *x*. Worldlines for light are lightlines. Worldline *OP* represents anything traveling with moderate constant speed compared to light speed (the slope gives a measure of the speed); *OQ* represents anything (except light) traveling with constant speed close to that of light; *IJ* represents events happening simultaneously; The *ct-axis* itself represents anything standing still at point *O (x,ct) = (0,ct)*; The *x-axis* represents events happening at *(x,ct) = (x,0)*.

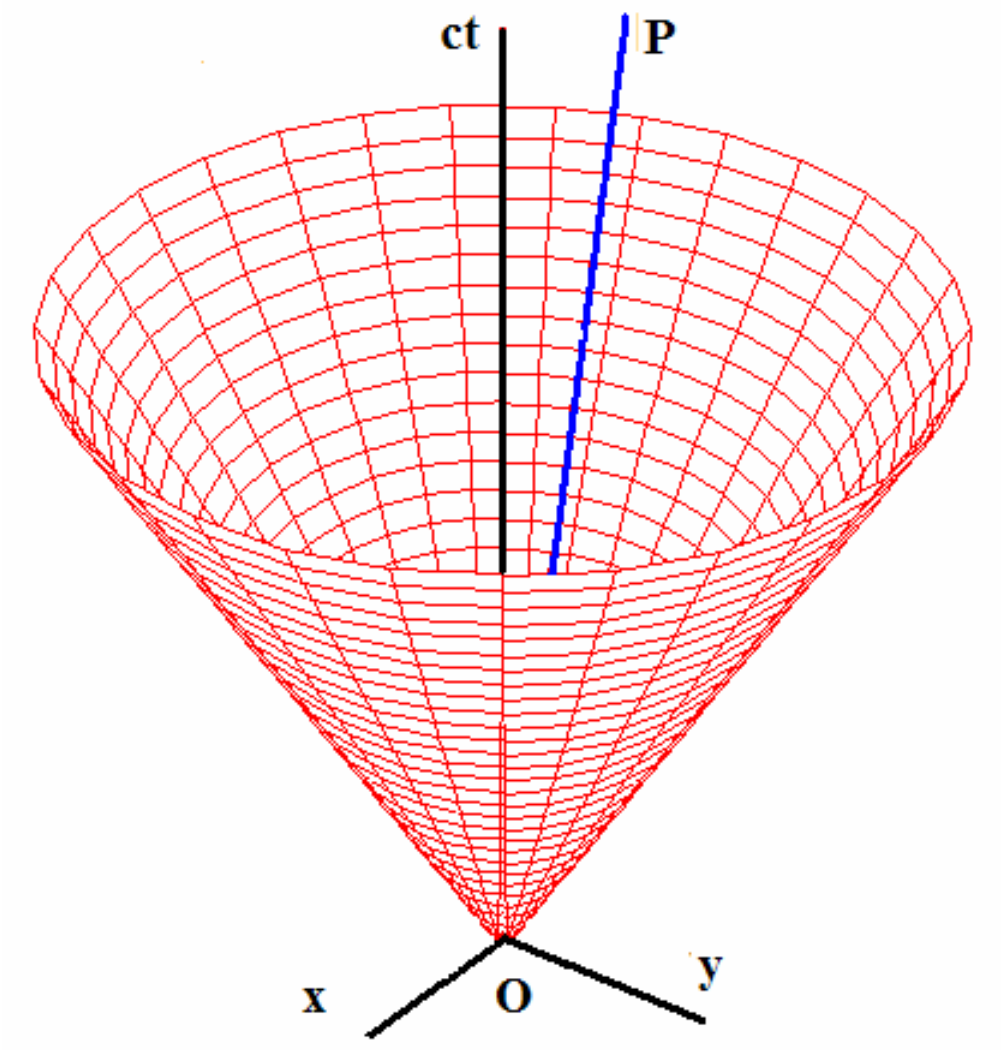

**Figure 2** – This shows spacetime diagram in three dimensions (*x,y,ct*). The lightlines form a light cone. The circles centered at *ct-axis* define simultaneity planes. The four dimensions case is impossible to be represented on a piece of paper. However most of the interesting effects between events in four-dimensional spacetime are reasonably represented in two and three dimensional spacetimes.

Let two inertial reference frames named $K = K(x,ct)$ and $K' = K'(x',ct')$ have rectilinear uniform relative motion, with speed $V$. For simplicity we will keep considering two-dimensional spacetime only.

Suppose now that you, the reader, are in the inertial frame *K*, at its spatial origin $(x,ct) = (0,ct)$ and some friend of yours is in *K'*, also at its origin $(x',ct') = (0,ct')$ which moves away with speed *V*, along *x-axis*. You may consider yourself to be in a "stationary" frame while your friend is in "non-stationary" frame. Since *K* and *K'* are both inertial frames, your friend may also, by himself, consider to be in a "stationary" frame, and judge that you are in a "non-stationary" inertial frame.

Assume that both *K* and *K'* are coincident at $(x,ct) = (0,0) = (x',ct')$.

As you have seen in Figure 1, the line which represents the light ray going to right, from the origin, bisects the angle between *ct-axis* and *x-axis*. In Figure 3 both *K* and *K'* are drawn according to the viewpoint of the observer in *K* (you). An arbitrary point event *E* is also represented in the diagram and has coordinates $(x_E,ct_E)$ in *K* and $(x'_E,ct'_E)$ in *K'*. The angle $\theta$, shown in the Figure 3, indicates the speed of *K'* relative to *K* ($tan(\theta) = V/c$). Such an event *E* and any other event with the same time coordinate $ct_E$ are considered simultaneous in *K*. Event *E* and any other event with the same coordinate $ct'_E$ are simultaneous in *K'*. So simultaneity is not an absolute concept, on the contrary, simultaneity is relative; the judgments of simultaneity of events will vary according to the state of motion of the observer. We should mention also that all the clocks at rest in a given inertial frame should be considered synchronized [1].

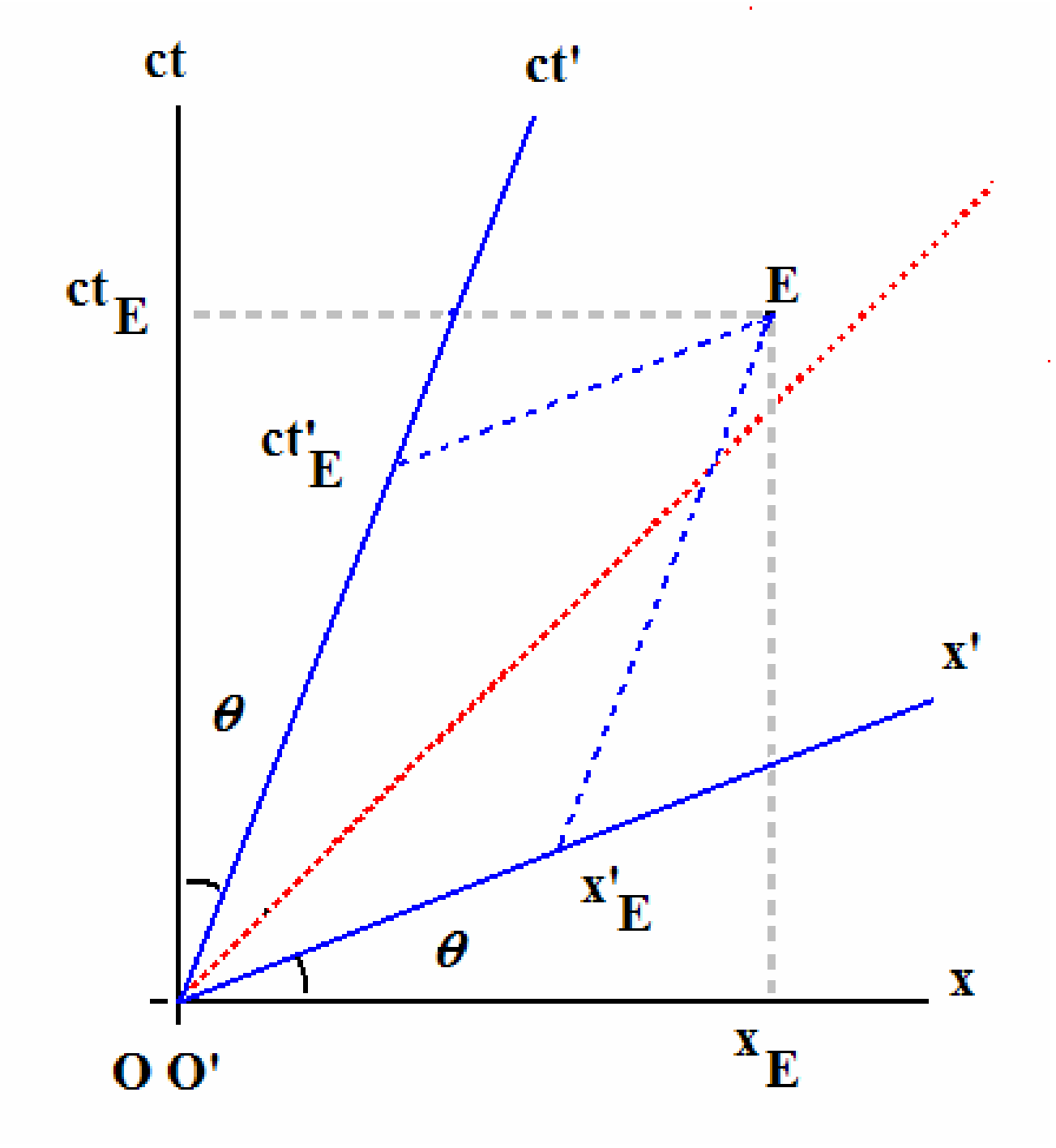

**Figure 3** – Both inertial reference frames *K* and *K'* are shown. As "seen" in frame *K*, the event *E* has $ct_E$ and $x_E$ coordinates. As "seen" in frame *K'* the event *E* has $ct'_E$ and $x'_E$ coordinates; $tan(\theta) = V/c$, where *V* is the speed with which *K'* is receding from *K*.

At this point you might be questioning if Minkowski spacetime is Euclidean. It is not. In two-dimensional spacetime diagram, a Minkowski space is represented on a Euclidean plane (sheet of paper) because the points in the Euclidean plane (events in spacetime) are labeled by pairs of real numbers (one for space and one for time).

Since Minkowski spacetime is not Euclidean, you cannot use the Pythagorean theorem in its familiar (Euclidean) form to calculate physical properties from Minkowski diagram. Later on in this paper you will find the substitute for such theorem.

All the events inside the lightcone centered at *O* belong to the *future* of *O* (*ct>0* and $|V|<c$). Obviously the lightcone has its counterpart on the negative side of *ct-axis* (ct<0 and |V|<c); it embodies all events which belong to the *past* of *O*. All region outside the lightcone is inaccessible to *O* ($|V|>c$). You may call it *elsewhere.*

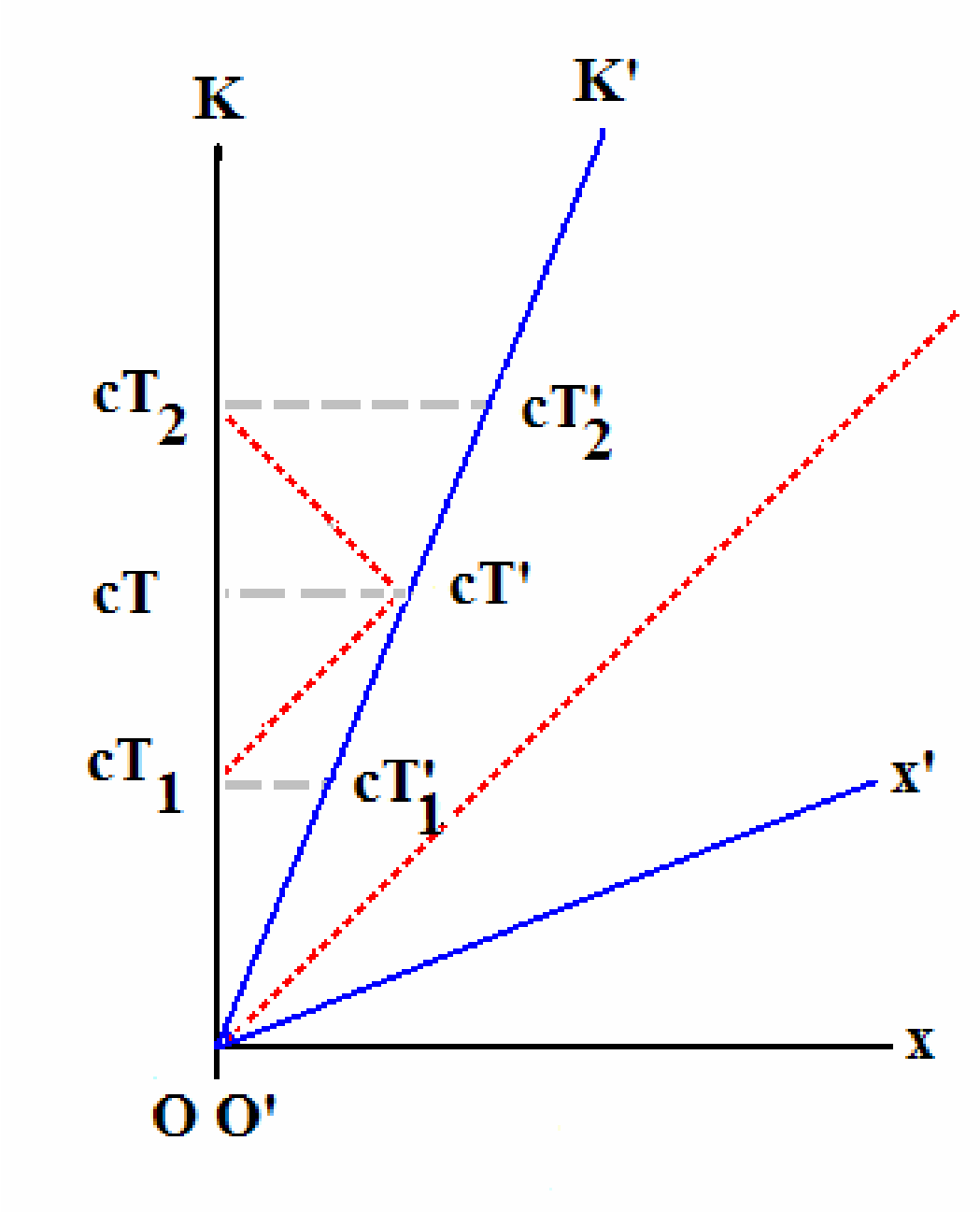

**Figure 4** – Here the inertial frame *K'*, moving with constant speed $V<c$ to the right, is represented as "seen" in *K*. The initial conditions are (*x,ct*) = (*0,0*) = (*x',ct'*). A light source at the spatial origin in K emits a pulse towards *K'* where exists a mirror that reflects it back. Your clock says that the pulse is emitted at event (0, $cT_1$) and is back at event (*0*, $cT_2$). According the postulate of constant speed of light one can get that $cT$ = ($cT2+cT1$)/2. Events (*0,cT*) and (*0,cT'*) are simultaneous in *K*.

## II - Doppler Factor

Consider now the Figure 4 in which *K'*, moving with constant speed *V* to the right, is represented as "seen" in *K*. The initial conditions are the same as before (*x,ct*) = (*0,0*) = (*x',ct'*). Suppose you (in *K*) have a light source at the origin which emits a pulse towards K' (your friend) where exists a mirror that reflects it back. Your clock says that the pulse is emitted at event (0, $cT_1$) and is back at event (*0*, $cT_2$). From Figure 4 you can get that the event *(0,cT)* where

$$cT = \frac{(cT_1 + cT_2)}{2} \tag{1}$$

is simultaneous to the event (*0,cT'* ), in *K'*, at which the pulse is reflected. It worth mentioning that *cT'* is

$$cT' = \frac{(cT_1' + cT_2')}{2}. \tag{2}$$

Knowing $cT$, you can calculate the spatial distance your friend was from you at the reflection event:

$$x = VT = \beta\, cT\,, \qquad (3)$$

where $\beta = \dfrac{V}{c}$. This process which gives you distance is the so called radar procedure.

Notice now from the Figure 4 that you may write that

$$cT_2 = cT + VT = (1+\beta)\, cT\,, \qquad (4)$$

and

$$cT_1 = cT - VT = (1-\beta)\, cT\,. \qquad (5)$$

Dividing expression (4) by expression (5) leads to

$$\frac{cT_2}{cT_1} = \frac{(1+\beta)}{(1-\beta)}\,. \qquad (6)$$

The event $(0, cT_1)$ at which the light ray is emitted is arbitrary; the ratios $cT_2/cT'$ and $cT'/cT_1$ should be the same and dependent only on $V$. Let $k$ be

$$k = k(V) = \frac{cT_2}{cT'} = \frac{cT'}{cT_1}\,. \qquad (7)$$

From expression (7) you get the very important result

$$cT' = \sqrt{T_1 T_2} \qquad (8)$$

which we'll be talking about a little later.

From expressions (8), (6) and (5) one has

$$k = \sqrt{\frac{1+\beta}{1-\beta}}\,. \qquad (9)$$

This is Doppler k-factor [**]. Notice the property

$$k(-V)\, k(V) = 1\,.$$

---

[**] Interesting to notice that the relative speed between the two frames is obtainable through the k-factor: $\beta = (k^2 - 1)/(k^2 + 1)$

## III - Time Dilation

The light ray that leaves the origin of your frame *K* at event (0, $cT_1$) hits the mirror at ($x=\beta\, cT,\ cT$), according to *K* . For your friend (in *K'*) such event occurs at ($x'=0, cT'$). We want to know the relation between *cT* and *cT'*, elapsed time since the common origin according to *K* and *K'*, respectively. From expressions (4) and (7) it follows that

$$\frac{cT_2}{cT'} = \frac{(1+\beta)\, cT}{cT'} = \frac{1}{k}, \qquad (10)$$

$$cT = \gamma\, cT' \qquad (11)$$

where $\gamma = \dfrac{1}{\sqrt{1-\beta^2}}$. (12)

Since $\gamma > 1$ you conclude that $cT > cT'$ which means that time flows slower for *K'*.

The elapsed time measured by a clock at rest is called a proper time. Then time *cT'* measured by the clock in *K'* at the same spatial position ($x'=0$) is called a proper time.

## IV - Length Contraction

Let a rod be at rest in *K* and let its length to be known by applying the standard measuring-rod as many times as necessary. Let such length be $L_0$ - the proper length - according to your inertial frame *K*. The proper length is the measured length of anything which is at rest in a given inertial frame. Let a light source be placed at the left end of the rod and a mirror at the right end. The mirror is placed in such way that it reflects back any ray coming from the other end of the rod. The rod is at rest in *K* and its left end coincides with the origin, and you have a standard clock at the light source position, so you are able to measure the time a light ray spends along the rod to go and come back to the starting point position. Suppose you measure the elapsed time and call it $2T_0$. Since the speed of light is constant no matter the direction it goes, according to Einstein's principle, you may say that $L_0 = cT_0$.

Now suppose this same rod appears to be at rest in *K'*. The rod length, as measured by your friend in *K'* in which it is at rest, is $L'_0$. As before the rod has the left end at $(x',ct') = (0,ct')$ and the right end at $(x',ct') = (L'_0,\ ct')$.

In Figure 5 the rod which is along *x-axis* (and *x'-axis*) is represented by the double line segment with length labeled $L'_0$ between *O'* and *A* points and is seen to be at rest in *K'*. The light ray emitted at the common origin of both frames moves along *x-axis* (and *x'-axis*) and reaches the right end of the rod at point event *Q* which has coordinates $(x,ct)=(L_{02}, cT_R)$ in your frame *K*. The *spatial distance* traveled by light to reach the right end of the rod is $OB = ct_R$, as measured in K,

$$cT_R = L + VT_R \qquad (12)$$

where $L$ is the presumed length of the rod as "seen" by you, in $K$, and you want to know its value:

$$L = (1-\beta)\, cT_R \,. \qquad (13)$$

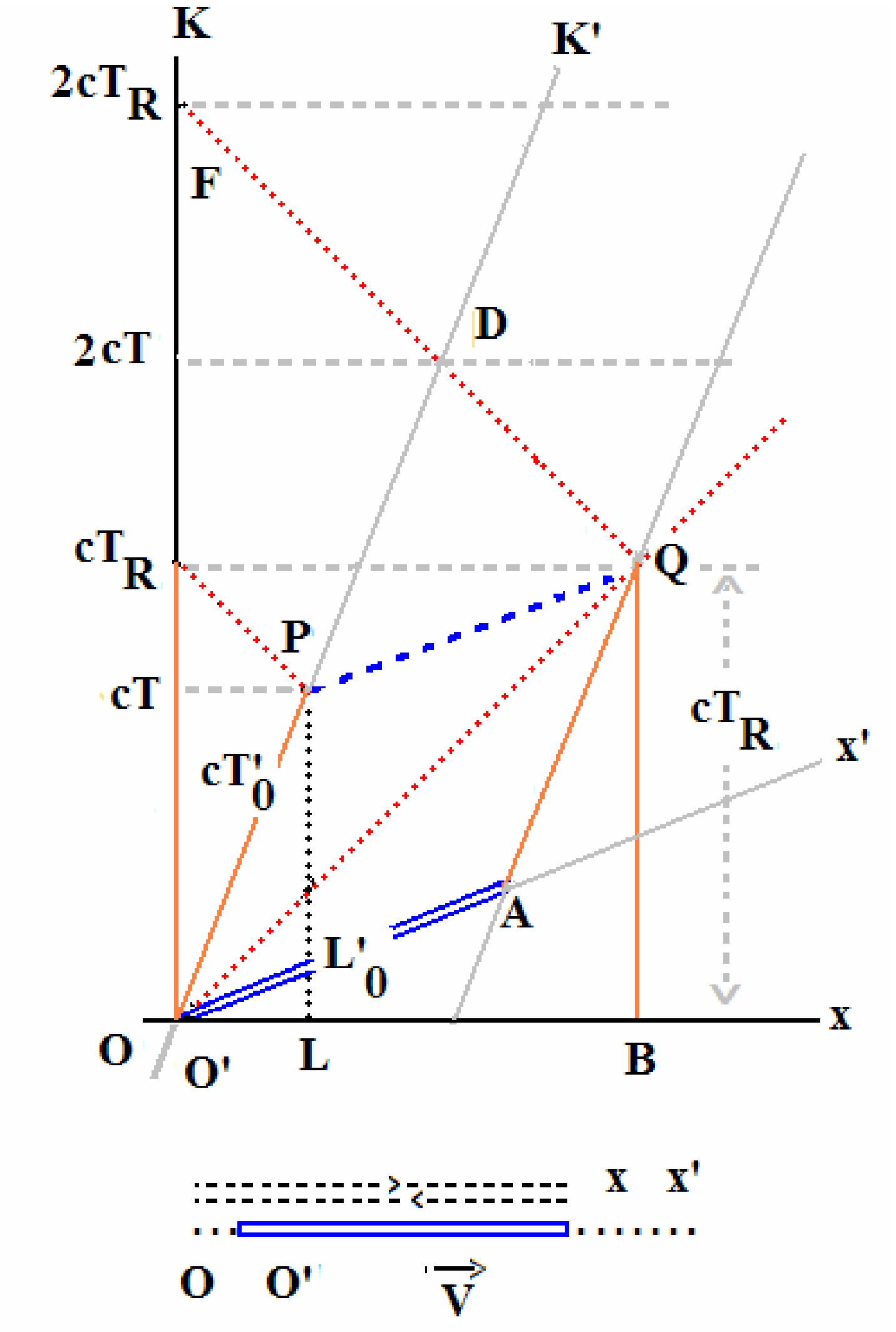

**Figure 5** – Inertial frame $K'$, moving with constant speed $V<c$ to the right, is represented as "seen" in $K$. The initial conditions are $(x,ct) = (0,0) = (x',ct')$. A rod which is along *x-axis* (and *x'-axis*) is represented by the double line segment with length labeled $L'_0$ between $O'$ and $A$ points and is seen to be at rest in $K'$. Its left end is at $(x',ct') = (0,ct')$ and the right end at $(x',ct') = (L'_0,\ ct')$. A light ray emitted at the common origin of both frames moves along *x-axis* (and *x'-axis*) and reaches the right end of the rod at event $Q$ $(x,ct)=(OB,cT_R)$, according to $K$.

According to your friend (in $K'$) the time the light ray spent from the left end to the right end of the rod is $cT'_0$. From equation (10) you calculate that

$$cT = \gamma\, cT_0' \qquad (14)$$

is the time you judge that the mirror at the right end of the rod reflects back the pulse. You do not have a direct measure of such time since your clock is at origin of $K$. But you can have a direct measure of the full back and forth time light spent ( $2cT_R$). Then according to the radar procedure, $cT_R$ is the time (in K) at which light reaches the mirror at the right end of the rod. Then, see Figure 4,

$$cT_R = \sqrt{\frac{1+\beta}{1-\beta}}\ cT_0' \ , \qquad (15)$$

$$L = (1-\beta)\, cT_R = (1-\beta)\sqrt{\frac{1+\beta}{1-\beta}}\ cT_0' = \sqrt{1-\beta^2}\, cT_0' = \frac{cT_0'}{\gamma} \qquad (16)$$

Since $L_0' = cT_0'$ is the proper length of the rod in *K'* then you are led to believe that

$$L = \frac{1}{\gamma} L_0' \qquad (17)$$

is the length the moving rod appears to have ($L<L_0$). This is the length contraction effect predicted by special relativity.

In a recent paper [3], one of us used two rods to prove the relativistic length contraction (expression 17). The scenario is similar to the one shown in Figure 5: one rod has proper length $L_{01}$ (equal to ours $L_0$) which is moving to the right with speed $V$ and a second rod, at rest, with proper length $L_{02}$. A light ray is emitted at the origin when both left ends coincide. Such light ray is expected to meet coincidently both rod's right ends. This triple meeting only occurs if $L_{02}$ is equal to distance *OB* in our Figure 5. That means $L_{02} = k\, L_{01}$. So one sentence is missing in that paper, such as "The proper length of both rods should be such that their right ends and the light ray meet at the same point in time, which means $L_{02} = k\, L_{01}$ where $k = ((1+V/c)/(1-V/c))^{1/2}$ is Doppler factor."

## V - Addition of Velocities

It is interesting to notice that the Doppler factor enables us to derive the addition of velocities expression in a quite straight work. Let three inertial frames $K$, $K_1$ and $K_2$, which clocks were all synchronized at the common origin. Let $V_1$ and $V_2$ be the velocities of $K_1$ and $K_2$ with respect to $K$.

From the figure and the definition of Doppler k-factor we can write

$$\frac{ct_1}{ct} = k_1(V_1) = \sqrt{\frac{1+\beta_1}{1-\beta_1}} \qquad \beta_1 = \frac{V_1}{c} \qquad (18)$$

$$\frac{ct_2}{ct} = k_2(V_2) = \sqrt{\frac{1+\beta_2}{1-\beta_2}} \qquad \beta_2 = \frac{V_2}{c} \qquad (19)$$

One then can ask what should be the *k-factor* for $K_2$ with respect to $K_1$. This shoul be

$$\frac{ct_2}{ct_1} = k_{12}(V_{12}) = \sqrt{\frac{1+\beta_{12}}{1-\beta_{12}}} \qquad \beta_{12} = \frac{V_{12}}{c} \qquad (20)$$

where $V_{12}$ should be velocity of $K_2$ with respect $K_1$.

From (18), (19) and (20) we get

$$k_2(V_2) = k_1(V_1)k_{12}(V_{12}) = \sqrt{\frac{1+\beta_1}{1-\beta_1}}\sqrt{\frac{1+\beta_{12}}{1-\beta_{12}}} = \sqrt{\frac{1+\beta_2}{1-\beta_2}} \qquad \beta_{12} = \frac{V_{12}}{c} \qquad (21)$$

which leads to the well known result for $V_{12}$

$$V_{12} = \frac{V_2 - V_1}{1 - \frac{V_1 V_2}{c^2}} . \qquad (22)$$

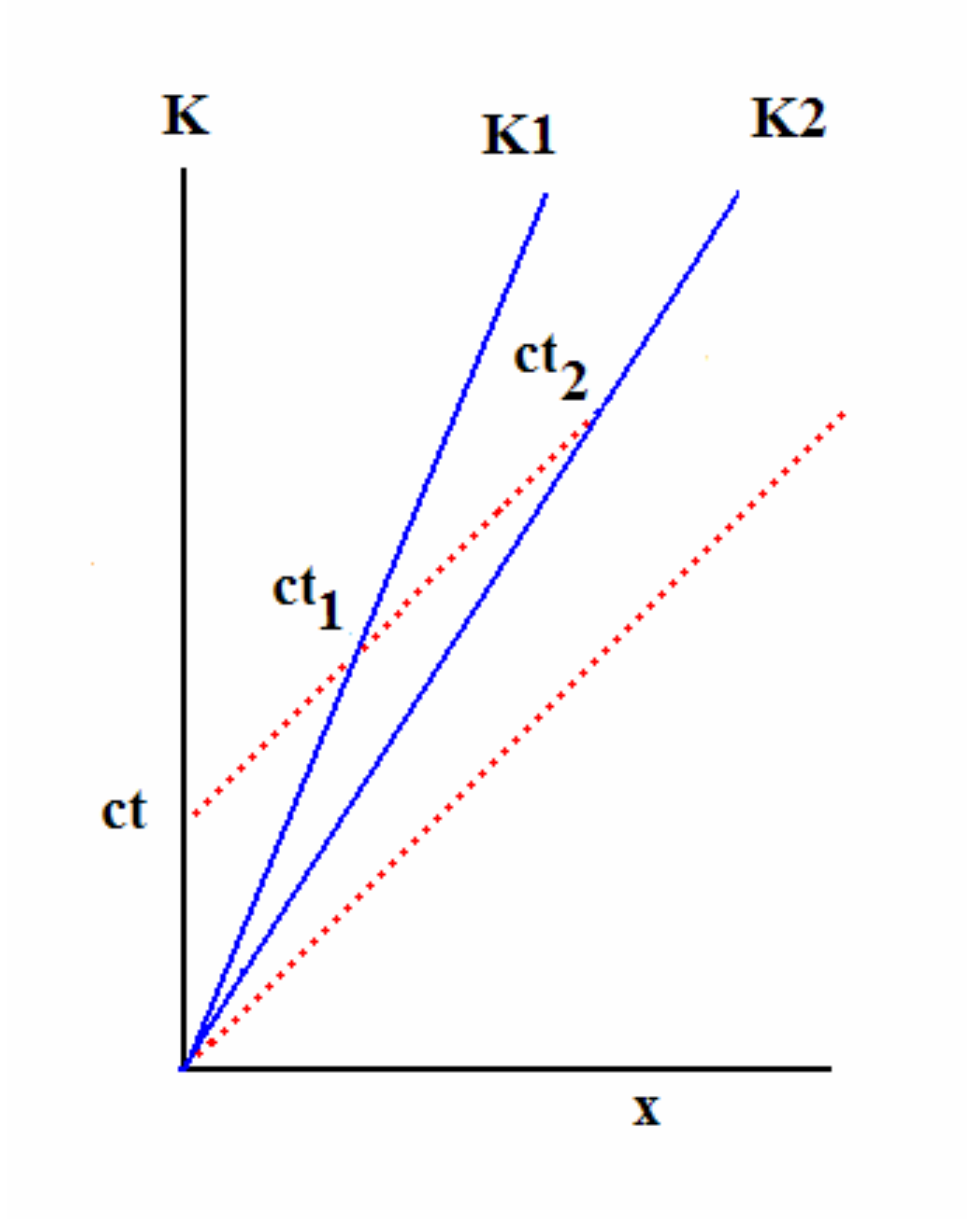

**Figure 6** – $K_1$ and $K_2$ are inertial reference frames traveling along *x-axis* with speeds (velocities) $V_1$ and $V_2$ with respect to $K$. Imagine a light ray emitted in $K$ at $(0,ct)$ which reaches $K_1$ at $(0,ct_1)$ and $K_2$ at $(0,ct_2)$. Knowing that $(ct_1/ct) = k_1$ and $(ct_2/ct) = k_2$ one gets $(ct_2/ct_1) = k_{12} = k_1k_2$, from which follows the expression for addition of speeds or velocities.

## VI - The Invariant Spacetime Interval

From what you have seen in section II and the Figure 7 you can write

$$\begin{aligned} ct_2 &= ct + x \qquad & ct_2' &= ct' + x' \\ ct_1 &= ct - x \qquad & ct_1' &= ct' - x' \end{aligned} \qquad (23)$$

$$\frac{ct_1'}{ct_1} = \frac{ct_2}{ct_2'} = k \qquad (24)$$

$$c^2 t_1 t_2 = c^2 t_1' t_2' \qquad (25)$$

From expressions (23) and (25) we get

$$(c^2t^2 - x^2) = (c^2t'^2 - x'^2) = s^2 \qquad (26)$$

where *s* does not depend upon which inertial frame one is dealing with, independs on the observer, it is invariant.

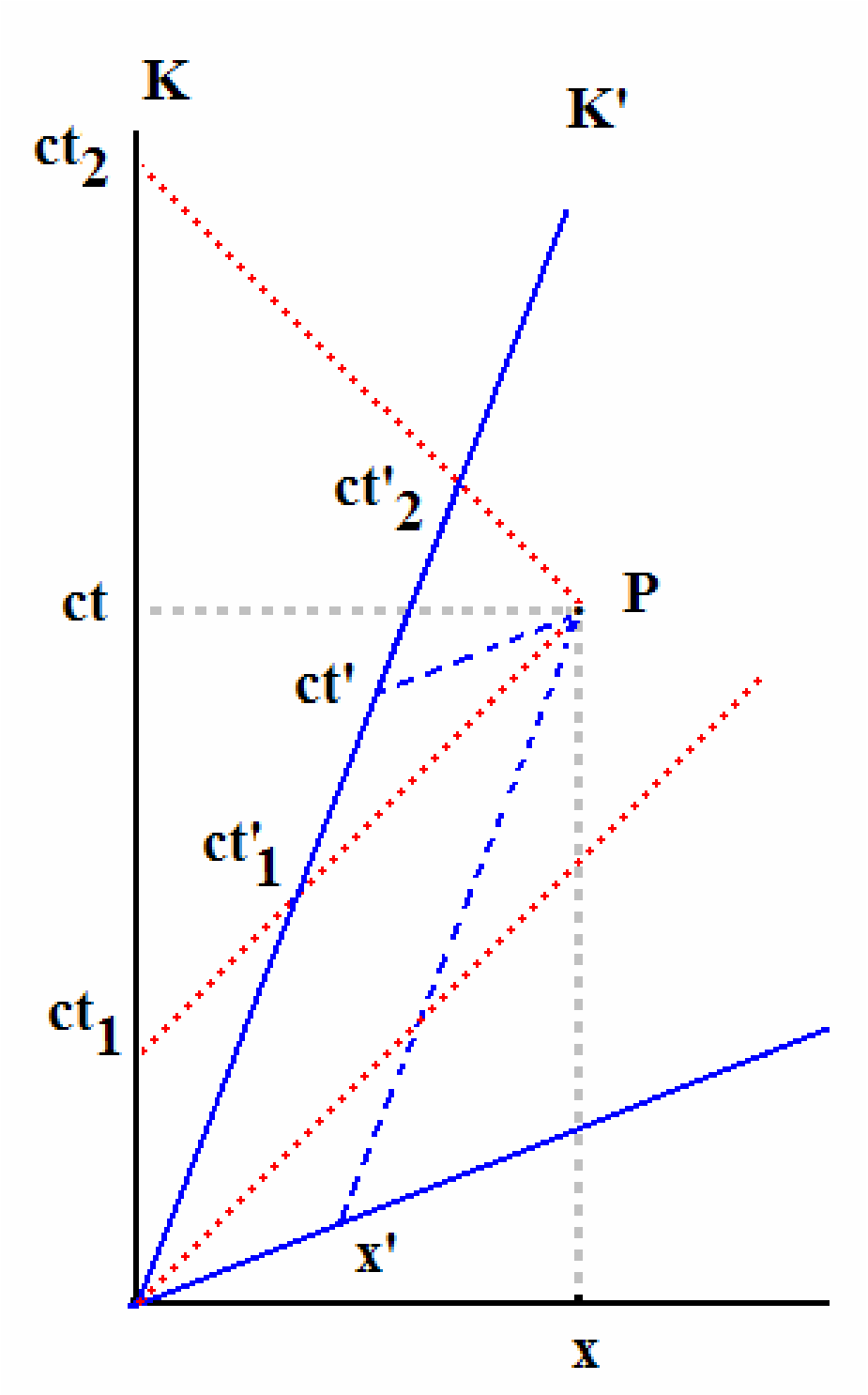

**Figure 7** – A light ray is emitted at *(0,ct₁)* in *K*, reaches *K'* at *(0,ct₁')* and then is reflected back at *(x,ct)* according to *K*. In returnig to the spatial origin in *K* it crosses *K'* at *(0,ct₂')* and then hits back to *K* at *(0,ct₂)*; *ct* is the midpoint between $ct_1$ and $ct_2$, and is simultaneous with *P* in *K*; *ct'* is also the midpoint between $ct_1'$ and $ct_2'$ and is simultaneous to *P* in *K'*.

It should be clear that when one talks about coordinates of a given event *E* in spacetime one actually is dealing with two events: event *E = (x,ct)* and event origin *O = (0,0)*. Very commonly one considers the observer at rest at the origin, or, the origin at the observer. Then one can write, with no loss of generality that

$$(c^2\Delta t^2 - \Delta x^2) = c^2\Delta t'^2 - \Delta x'^2 = \Delta s^2 \qquad (27)$$

The quantity $\Delta s$ is called *spacetime interval* and it is an invariant in special relativity. If one were talking about Euclidean space this would play the rolle of distance between two points. If spacetime were Euclidean one *would* calculate such interval as $(c^2\Delta t^2 + \Delta x^2)$, by using the Pythagorean theorem. So here there is a critical difference between Euclidean space and spacetime: "*Pythagorean thorem*" is much different here.

There is a good paradigm for a proper time, which is the time your wristwatch shows, assuming of course that it is at rest relative to you. It is common among the physicists to refer to proper time by the greek letter $\tau$. Then for $\Delta x' = 0$, $\Delta\tau = \Delta t'$ you have:

$$(c^2\Delta t^2 - \Delta x^2) = c^2\Delta\tau^2 = \Delta s^2 \qquad (28)$$

or

$$(c^2t^2 - x^2) = c^2\tau^2 = s^2 \qquad (29)$$

and if you are dealing with infinitesimals, you have the spacetime line element *ds*

$$(c^2dt^2 - dx^2) = c^2d\tau^2 = ds^2\,. \qquad (30)$$

You can easily see that $\Delta s^2 = c^2\Delta\tau^2$ (or $ds^2 = c^2d\tau^2$) can be greater than zero, equal to zero or less than zero. In the first case ($\Delta s^2 > 0$) the spacetime interval is said to be "timelike"; in the second case ($\Delta s^2 = 0$) it is "lightlike"; the last case ($\Delta s^2 < 0$) the spacetime interval is said to be "spacelike". This spacetime interval is an important quantity in further studies in relativity.

## VII - Lorentz Transformations

From the spacetime diagram you straightforwardly derive the Lorentz Transformations. Playing with expressions (23) and (24) you obtain

$$2x' = (ct_2' - ct_1') = (\frac{1}{k}ct_2 - kct_1) = (\frac{1}{k} - k)ct + (\frac{1}{k} + k)\,x \qquad (31)$$

$$2\,ct' = (ct_2' + ct_1') = (\frac{1}{k}ct_2 + k\,ct_1) = (\frac{1}{k} + k)\,ct + (\frac{1}{k} - k)\,x \qquad (32)$$

which with the expression for Dopler *k-factor* leads to Lorentz Transformations

$$\begin{aligned} x' &= \gamma\,(x - \beta\,ct) \\ ct' &= \gamma\,(ct - \beta\,x) \end{aligned} \qquad (33)$$

The inverse transformations are obtained by changing $\beta$ for $-\beta$ and primed for unprimed:

$$\begin{aligned} x &= \gamma\,(x' + \beta\,ct') \\ ct &= \gamma\,(ct' + \beta\,x') \end{aligned} \qquad (34)$$

## VIII – Calibrating Minkowski Diagram

At this point one should talk about calibration of spacetime diagram. The *ct-axis* and *x-axis* in all diagrams we have been using in this paper are already calibrated, as one tacitly assumed; the units of *ct* and *x* are assumed to be the same as it naturally results from the choice of coordinates (*x*,*ct*). They both have dimensions of distance. This actually is a reflex of the interdependence of space and time.

If spacetime were Euclidean one *would* draw a diagram as the one shown in Figure 8 and the calibration would be done through the use of circles ($c^2t^2 + x^2 = n^2$), centered at the origin, where $n$ are integer numbers. The calibrating circles define the locus of all points whose distance from the origin is multiple of the unit.

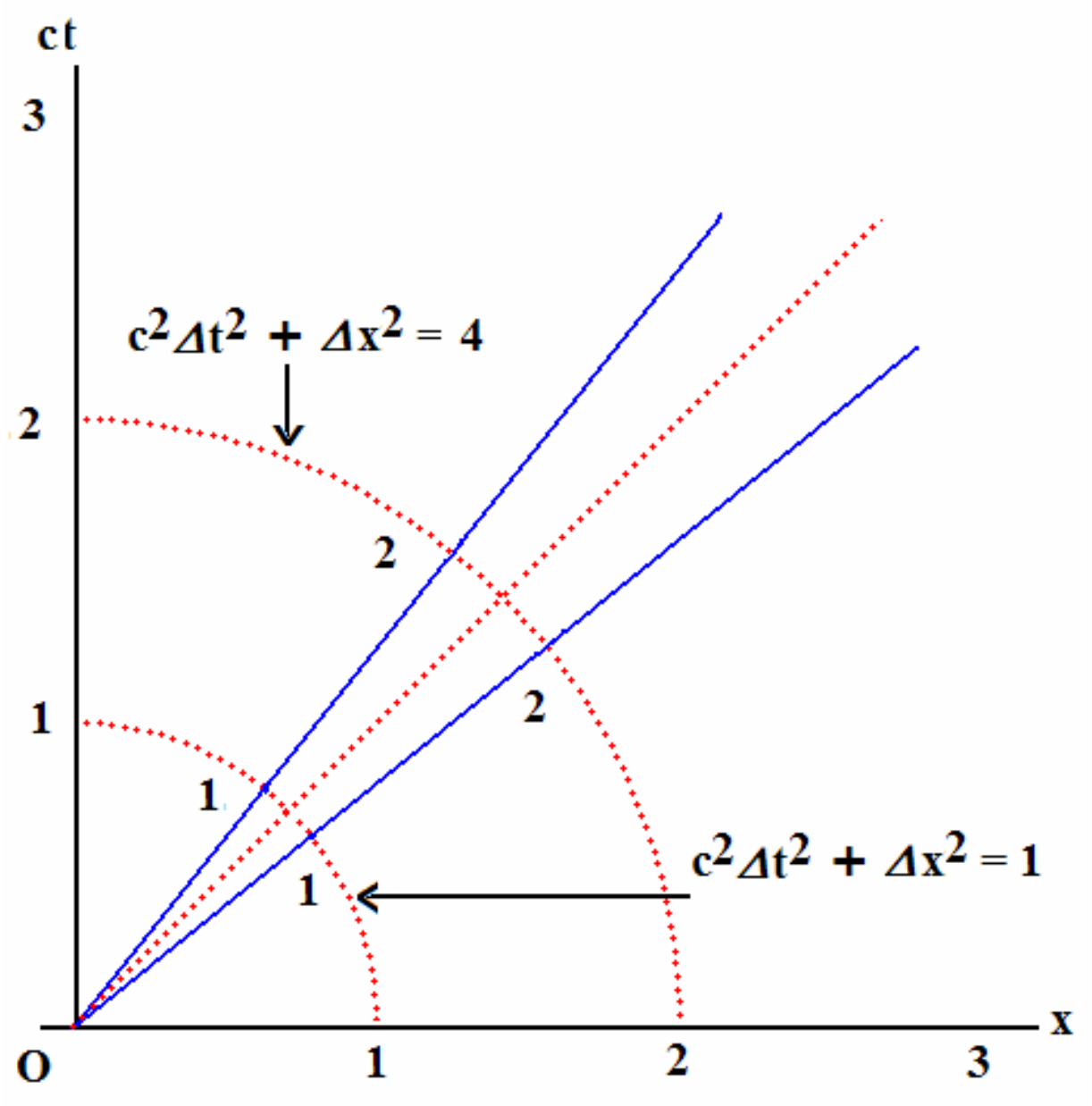

**Figure 8** – Spacetime is not Euclidean, but if it were one *would* draw a diagram as the one shown here and the calibration would be done through the use of circles ($c^2t^2 + x^2 = n^2$), centered at the origin, where $n$ are integer numbers. The calibrating circles define the locus of all points whose distance from the origin is multiple of the unit

Spacetime is not Euclidean and in order to do the calibration one has to use calibrating hyperbolas ($c^2t^2 - x^2 = \pm n^2$), with $n$ being integer numbers. You have seen that $\Delta s^2 = c^2\Delta t^2$ can be positive, null or positive, and this explains the ± signs in the hyperbola equations. In Figure 9 you can see the spacetime diagram with the calibrating hyperbolas (two branches for $n = 1$ and two branches for $n = 2$).

Let us consider the case for which $\Delta s^2 = +1$. The *1* on *ct-axis* defines the event $(x,ct) = (0,1)$ in reference frame $K$ (you). Along the same hyperbola ($\Delta s^2 = +1$), the *1* on *ct'-axis* defines the event $(x',ct') = (0,1)$ in $K'$ (your friend). You can promptly see that while the time is $\Delta t = 1$, in accord to $K$, in $K'$ the time should be $\Delta t' < 1$. This just the time dilation phenomenon: $\Delta t = 1 = \gamma\,\Delta t'$, $\Delta t' = 1/\gamma < 1$. So time flows slower for $K'$, under the viewpoint of $K$. A symmetrical situation occurs for the observer in $K'$. To him while his clock is showing $\Delta t' = 1$, following K' simultaneity line, you would find out that your clock shows $\Delta t < 1$. This just the same time dilation phenomenon as "seen" by the observer in $K'$; $\Delta t' = 1 = \gamma\,\Delta t$, $\Delta t = 1/\gamma < 1$. So time flows slower for $K$, which is receding with speed $-V$ under the viewpoint of $K'$.

Now the case for which $\Delta s^2 = -1$. The *1* on *x-axis* defines the event $(x,ct) = (1,0)$ in reference frame $K$ (you). Along the same hyperbola ($\Delta s^2 = -1$), the *1* on *x'-axis* defines the event $(1,0)$ in $K'$ (your friend). You can promptly see that the length $\Delta x = 1$ (at $\Delta t = 0$) in $K$ should be, according to $K'$, $\Delta x' = 1/\gamma < 1$. So a unit length in frame $K$ is "seen" as less than unit length in frame $K'$. This is the phenomenon of length contraction. Again here a symmetrical situation occurs. The *1* on *x'-axis* defines the event $(x',ct) = (1,0)$ in reference frame $K'$ (your friend). Along the same hyperbola ($\Delta s^2 = -1$), the *1* on *x-axis* defines the event $(1,0)$ in $K$ (you). While $\Delta x' = 1$ in $K'$, you find out that the same length in $K$ should be smaller ($\Delta x = 1/\gamma < 1$); just follow the equilocality line (same spatial position according to $K'$) starting from $(x',ct') = (1,ct')$.

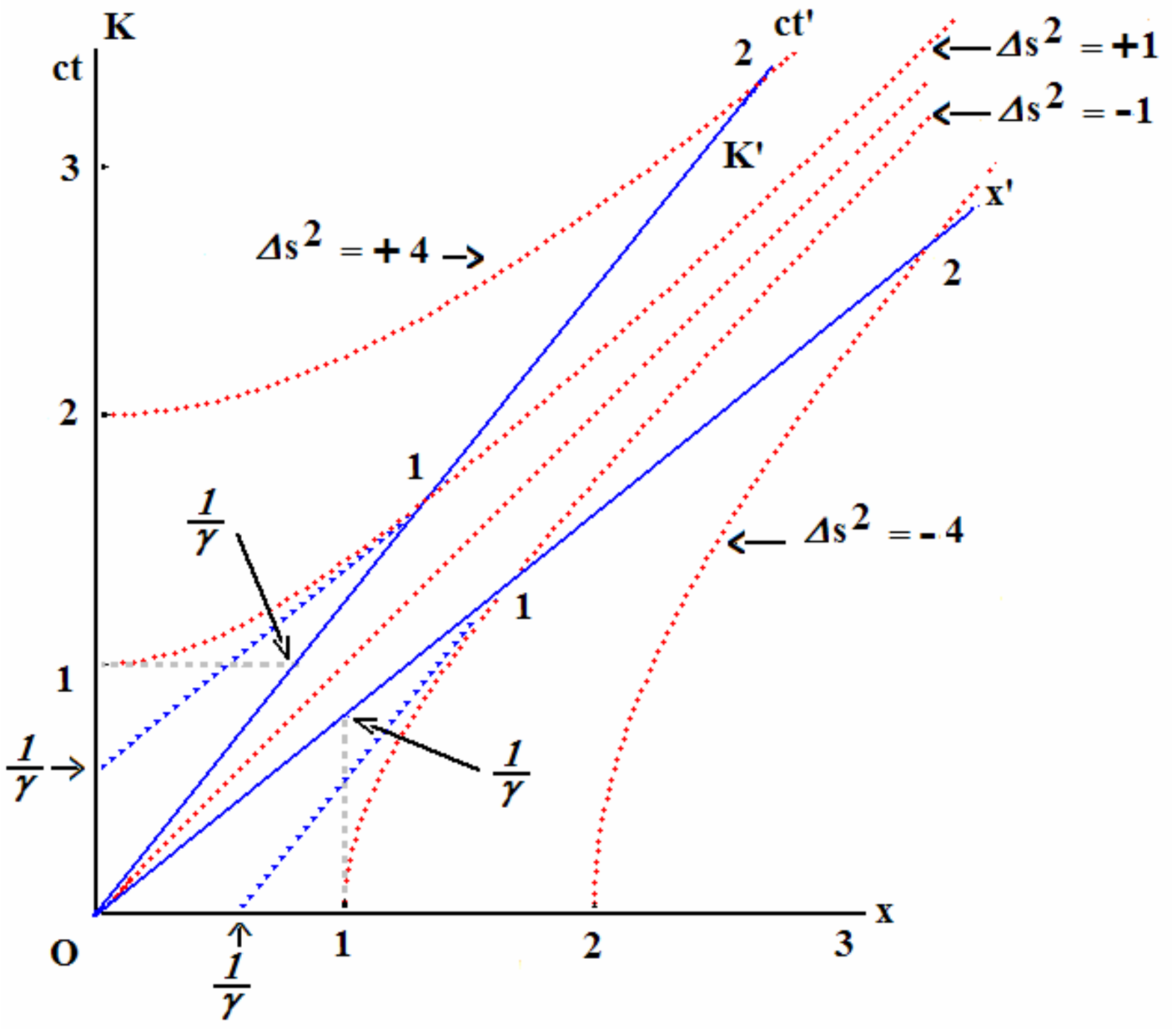

**Figure 9** – Spacetime diagram is calibrated by using calibrating hyperbolas ($c^2t^2 - x^2 = \pm n^2$), with $n$ being integer numbers. Here one draws two hyperbolas (for $n = 1$ and $n = 2$). One can see quite clear, at least qualitatively, the time dilation and the length contraction properties and their symmetrical aspects.

## Conclusion

Starting with Minkowski diagram we get Doppler k-factor, time dilation, length contraction, addition of velocities, invariant spacetime interval and Lorentz Transformations. Hope this approach may be useful in complementing standard textbooks on special relativity.